\documentstyle[epsf,here,apalike,12pt]{article}
\input epsf
\newcommand{\ddGt}{\Delta \Delta G_{\ddagger-D}}
\newcommand{\ddGn}{\Delta \Delta G_{N-D}}
\newcommand{\med}{\mbox{med}}

\begin{document}
\title{ \bf Evolutionary conservation of the folding nucleus }
\author{Leonid Mirny and Eugene Shakhnovich}
\date{Nov 21, 2000}
\maketitle
\begin{center}
\vspace{1.5in}

Running title: Conservation of folding nucleus

\vspace{0.5in}
Submitted to {\it Journal of Molecular Biology}
\vspace{0.5in}

Harvard University, Department of \\
Chemistry and Chemical Biology \\
12 Oxford Street, Cambridge MA 02138 \\
\vspace{10pt}
{\sf E-mail: leonid@origami.harvard.edu, eugene@belok.harvard.edu \\
http://paradox.harvard.edu/\~ \ leonid}
\end{center}

\newpage
\begin{abstract}
In this Communication we present statistical analysis of conservation
profiles in families of homologous sequences for nine proteins whose
folding nucleus was determined by protein engineering methods. We show
that in all but one protein (AcP) folding nucleus residues are
significantly more conserved than the rest of the protein.  Two
aspects of our study are especially important: 1) grouping of amino
acids into classes according to their physical-chemical properties and
2) proper normalization of amino acid probabilities that reflects the
fact that evolutionary pressure to conserve some amino acid types may
itself affect concentration of various amino acid types in protein
families. Neglect of any of those two factors may make physical and
biological ``signals'' from conservation profiles disappear.

\end{abstract}

\section*{Introduction}

It is now widely accepted that folding of small single-domain proteins
follows ``nucleation-condensation'' mechanism
\cite{NUCLEUS,FER_NUCL,FER_COSB,COSB,GT,CO_PANDE} whereby relatively
small fragment of protein structure is formed in the transition state
between unfolded and folded states. Residues belonging to this
fragment constitute specific folding nucleus (SFN). Considerable
experimental \cite{FER_NUCL,Jackson_JMB,Serrano_NSB1,Chiti_NSB} and
theoretical \cite{NUCLEUS,KT_JMB,LEWYN_NSB,Dokh:2000} effort has been
devoted to identification of folding nuclei in real proteins and
various models as well as factors that determine its location in
structure and in sequence.

One of the most intriguing aspect of nucleation-condensation mechanism
of protein folding is its relation to protein evolution. Indeed
residues constituting folding nucleus can be metaphorically
considered ``accelerator pedals'' of folding \cite{EVOL2} since
mutations in those positions affect folding rate to a much greater
extent than elsewhere in a protein.  One can conclude that if there is
evolutionary control of folding rate it should have resulted in
additional pressure applied on folding nucleus residues, and such
pressure can be manifested in noticeable additional conservation of
nucleus residues.

This idea was first proposed in \cite{ALAMUT} where it was applied to
prediction of nucleus residues from protein structure. Many sequences
were designed to fit the structure of Chymotripsin Inhibitor 2 (CI2)
with low energy. Positions conserved among the designed sequences were
identified as a putative nucleus. This way {\em blind} predictions of
folding nucleus in CI2 were made that were verified in independent
experiments \cite{FER_NUCL}.  

 In related studies
papers Ptitsyn studied conservatism in distant yet related by sequence
homology members of Cytochrome C \cite{PTIT_EV} and myoglobin
\cite{PTIT_EV2} families.  In both cases he found conserved clusters
of residues without an obvious functional role which he suggested to belong to
folding nucleus of those proteins. Michnick and
Shakhnovich \cite{UBQ_EVOL} carried out an analysis of conservation
in natural and designed sequences 
for families of three structurally related proteins - ubiquitin, raf
and ferredoxin and predicted possible folding nucleus for those proteins.

Neverteheless the notion of folding nucleus conservation has drawn some
controvercy in the lietrature.  While earlier papers
\cite{ALAMUT,UBQ_EVOL,PTIT_EV,PTIT_EV2} suggested  conservation of folding
nucleus in some proteins, a more recent paper by Plaxco and coauthors
\cite{Plaxco:2000} argued to the opposite. These authors looked at
conservatism profile in several protein families for which protein
engineering analysis of folding transition states has been carried
out, and did not observe correlation between conservation and 
experimentally measured $\phi$-values.  This made them conclude that 
there is no evolutionary pressure to control the folding rates.

In this work  we study evolutionary conservation of the folding nucleus for
several homologous proteins. Conservation of the folding nucleus is
systematically compared with the conservation in the rest of the
protein sequence. In contrast to previous studies, we perform rigorous
statistical test to assess significance of higher conservation in the
folding nucleus.  The main result of this study is that for all
studied proteins, except AcP, folding nucleus is significantly more
conserved than the rest of the protein.  We explain the difference
between our thorough statistical analysis and that of Plaxco et al
\cite{Plaxco:2000} by pointing out to some technical shortcomings in the
earlier work \cite{Plaxco:2000}.

\section*{Results and Discussion}
To study evolutionary conservation of the folding nucleus we turn to
nine proteins for which nucleus has been experimentally identified
from protein engineering analysis: CI2, FKBP12, ACBP, CheY, Tenascin,
CD2.d1, U1A, AcP and ADA2h. For each of them we obtain a multiple
sequence alignment from HSSP database \cite{HSSP} (or PFAM \cite{pfam}
database if HSSP contains too few sequences). We compute variability
at position $l$ of the alignment as
\begin{equation}
s(l)= - \sum_{i=1}^{6} p_i(l) \log p_i(l)
\label{eq:s}
\end{equation}
where $p_i(l)$ is the frequency of residues from class $i$ in position
$l$. We use six classes of residues to reflect physical-chemical
properties of amino acids and their natural pattern of substitutions:
aliphatic [A V L I M C], aromatic [F W Y H], polar [S T N Q], basic [K
R], acidic [D E], and special (reflecting their special conformational
properties) [G P]. As a result of this classification mutations within
a class are ignored (e.g. $V \rightarrow L$), while mutations that
change the class are taken into account. Figure 1 presents variability
profile for studied proteins with nucleation positions marked by
filled circles. 
\marginpar{Fig.1}
 Importantly, we defined the folding nucleus as it was
identified by the original experimental groups (Table 1).

Figure 2  \marginpar{Fig.2} clearly shows that nucleus residues are almost 
always among the most
conserved ones for all studied proteins. It also shows that nucleus
residues are not the only conserved ones: many other residues
(predominantly in the cores of the proteins) are also conserved.

In order to evaluate statistical significance of nucleus conservation 
we 
 compare evolutionary conservation of the folding nucleus with the
conservation of all residues in the protein using the following
statistical test. We start from the null hypothesis H0 that nucleus
residues are {\bf no} more conserved than the whole protein
sequence. To test this hypothesis we compute median variability of
the nucleus residues ($\med[s_{nuc}]$) and compare it with the
distribution of medians variability of the same number of residues
randomly chosen in the same protein ($f(\med[s_{rand}])$). The
distribution $f(\med[s_{rand}])$ is obtained by choosing $10^5$ random
sets of $n$ residues ($n$ is the number of residues in the
nucleus). Then the fraction of instances with
$\med[s_{rand}]<\med[s_{nuc}]$ gives the probability $P_0$ of
accepting H0. In other words, $P_0$ is the probability that observed
lower variability of the folding nucleus is obtained by chance. Hence,
$P_0 \leq \alpha$ indicates statistically significant strong
evolutionary conservation of the folding nucleus.  Below we use
confidence level $\alpha=2\%$.

Table 2 presents computed $P_0$ values. {\it The main result of this
work is that in all proteins, except AcP, residues in the folding
nucleus are significantly more conserved than the rest of the
protein. }

Next we study how obtained results depend on the way amino acids are
grouped into classes (see Table 2). When classification scheme from
\cite{BT98} (BT) is used, still all proteins except AcP exhibit
significant conservation of the folding nucleus. This clearly
demonstrates that observed conservation of the folding nucleus is not
a consequence of a particular choice of the classification scheme.

However, when amino acids are {\em not} grouped into classes, nucleus
exhibits significant conservation only in four out of nine proteins.
Taken together these results indicate that substitutions in the
folding nucleus may occur, but they  are limited to
residues that belong to the same class (i.e. have similar
physical-chemical properties \cite{Thompson96}).

To study what physical-chemical properties are conserved in the
folding nucleus we used various classification schemes. Starting from
all 20 amino acids, we grouped some of them into classes and repeated
the analysis, including the statistical tests (see Table 2).  The goal
is to find a {\em minimal} classification (i.e. grouping the {\em
minimal} number of amino acids together) that provides statistically
significant conservation of the folding nucleus. Our results show that
classification where only I, L, and V are grouped in one class while
all other amino acids each represent their own class satisfies this
requirement (see Table 2).This classification provides significant
conservation of the nucleus for all proteins except AcP with
$\alpha=5\%$, and for all proteins except AcP and FKBP12 with
$\alpha=2\%$. This result demonstrates that $ I \rightleftharpoons L
\rightleftharpoons V$ are the most common substitutions in the nucleus
(and in the protein core in general \cite{BLOSUM62,GONNET}). These
substitutions are tolerated in the nucleus as they do not change much
neither stability of the native fold nor the folding rate. Analysis of available
experimental data (L.Li unpublished) shows that 
changes in stability  upon $ I
\rightleftharpoons L \rightleftharpoons V$ mutations are in average
$\langle \ddGn \rangle = 1.0 \pm 0.4 $kCal mol$^{-1}$ for the
native state and $\langle \ddGt \rangle = 0.2 \pm 0.3$ kCal
mol$^{-1}$ for the transition state.

Note that grouping of residues into classes to assess conservation is
similar to the use of substitution matrices in sequence alignment
techniques. The underlying idea for both methods is to take into
account natural physical-chemical similarity between amino acids and
their substitution patterns.  Plaxco et all used all 20 types of amino
acids and failed to identify strong conservation of the folding
nucleus \cite{Plaxco:2000}.  Similarly, a method that relies on simple
sequence identity cannot detect distant homology. However distant
homology between sequences can be detected using proper substitution
matrices \cite{Abagyan97,Brenner98}. The use of substitution matrices
is physically meaningful since they weight, e.g., $I-V$ match higher
then $I-D$, while a method that relays on percentage of sequence
identity weights $I-V$ and $I-D$ equally. Likewise, our amino acid
classification scheme does not count $I \rightarrow V$ as a mutation,
while it certainly considers substitutions like $I \rightarrow D$ as
mutations to be counted.

Although, on average, nucleus is more conserved, than the rest of the
protein, not all nucleating residues are strongly conserved. For
example, in CheY two out of ten nucleation residues are not
conserved. In ADA2h two out of five and in tenascin one out of four
residues are not conserved.  Some nucleus residues may be less
conserved because they belong to ``extended nucleus'' \cite{EVOL_JMB} or
because of limitation of our residues classification scheme that puts
aromatic and aliphatic residues into two different groups, while
aromatic-aliphatic substitutions may occur in the core of some
proteins (i.e. tenascin, ADA2h) usually as a result of correlated
mutations that are not treated properly in this approach (but are
taken into account in the conservation-of-conservation approach
\cite{EVOL_JMB}).  Another interesting observation is that the only protein
that exhibits no preferential conservation of the folding nucleus is
AcP, which is the slowest folding protein among all studied two-state
folding proteins ($k_f^{H_2O}=0.23 s^{-1}$). Perhaps, this protein did
not undergo evolutionary selection for faster folding and hence its
folding nucleus is under no additional pressure to be conserved.

Note that, as expected, several other residues in studied proteins are
as conserved as the nucleating ones. (see Fig.2) Those are the residues of the
active site, core hydrophobic residues responsible for stabilization
of the native structure and others. This suggests that although
folding nucleus is conserved it can not be uniquely identified just by
analysis of {\em a single} protein family as a pattern of conservation
is dominated by residues conserved for protein stability and function
(see \cite{MS_REVIEW}). Thus a consistent analysis should discriminate
between residues that are conserved for functional reasons, for
stability reasons and for kinetic reasons (folding nucleus), like it
was done in a more detailed conservation-of-conservation
analysis in \cite{EVOL_JMB}.

Why do results of our analysis differ from those of
Plaxco et al \cite{Plaxco:2000}? First, we took into 
account physical-chemical properties
of amino acids and their natural substitution patterns to group amino
acids into classes. As we showed, substitutions of large aliphatic
residues (I,L,V) are frequent in folding nuclei and this confused
previous analysis that did not apply any amino acid classification
scheme. While Plaxco et al claimed in their paper \cite{Plaxco:2000}
(without providing a supporting evidence) that grouping of amino acids
into classes did not change their conclusions, our analysis shows that
proper classification of amino acids is crucial for detecting
conservation in the folding nucleus.

Second, Plaxco et al used a different method to compute sequence
variability:
\begin{equation}
s_2(l)= - \sum_{i} p_i(l) \log [p_i(l)/p^0_i]
\label{eq:Plaxco}
\end{equation}
This equation differs from eq.(\ref{eq:s}), used in this study, in
normalization by $p^0_i$ - the ``background'' frequency of residue
type $i$ in all proteins. Although the difference may seem technical,
equations (\ref{eq:s}) and (\ref{eq:Plaxco}) are based on two
different models of evolution. We argue that while equation
\ref{eq:Plaxco} may be adequate for DNA sequence analysis
\cite{Stormo98} it is not appropriate for analysis of protein
evolution.

Equation \ref{eq:Plaxco} implicitly assumes that amino acid
composition $p^0_i$ is fixed {\em a priori} in each protein.  Hence
equation (\ref{eq:Plaxco}) tends to underestimate conservation of
``frequent'' amino acids (L,A,S etc), while overestimating
conservation of less frequent amino acids (W,C,H etc). In contrast,
equation (\ref{eq:s}) assumes that conservation requirement itself
affects the composition, i.e. higher conservation of an amino acid
leads to its higher frequency in proteins.

To illustrate this point consider a toy protein that consists of two
types of residues: hydrophobic H and polar P. Assume that $70\%$ of
amino acids in this proteins are in the core and $30\%$ are in the
loops. Also assume that in the toy world selection for stability
requires a 100\% conservation of H amino acids in the core, while loops
are under no evolutionary pressure and H and P are equally probable in
the loops. Then $p^0_H=1 \cdot 0.70+0.5 \cdot 0.3=0.85$ and
$p^0_P=0.5 \cdot 0.3 = 0.15$. At conserved core positions
$s_2(\mbox{core})= - 1 \log 1/0.85 \approx -0.16 $, while in the loops
$s_2(\mbox{loops})= -0.5 \log 0.5/0.85 - 0.5 \log 0.5/0.15 \approx
-0.34$. Hence, the use of equation (\ref{eq:Plaxco} leads to a 
counterintuitive and apparently 
wrong result
$s_2(\mbox{core})>s_2(\mbox{loops})$, i.e. that loops are more conserved
than 100\% conserved core! Clearly this result 
shows inadequacy of equation (\ref{eq:Plaxco})   
as applied to protein evolution
with unconstrained composition. Similarly, application of equation \ref{eq:Plaxco}  
to real proteins leads to unreasonably low conservation of the
hydrophobic core as compared to exposed loops (data not shown).

A possible  way to compensate for variations in amino acid 
composition of proteins is to
define the sequence entropy as in \cite{Schneider99}:
\begin{equation}
s(l)= - \sum_{i} p_i(l) \log p_i(l) + \sum_{i} p^0_i \log p^0_i
\label{eq:Sminus}
\end{equation}
where the second term gives the ``background'' variability due to
amino acid composition. This term however does not depend on $l$ and
hence does not change the relative variability.

Interestingly, the use of equation (\ref{eq:Plaxco}) by Plaxco et al
\cite{Plaxco:2000} gave rise to a surprising result that active sites
in proteins are generally no more conserved than the rest of the
protein (see Fig.2 of \cite{Plaxco:2000}).  Conservation of known active sites
 was used as a
control in \cite{Plaxco:2000} for their method of analysis based on
equation \ref{eq:Plaxco} which it apparently failed.

Finally, Plaxco et al did not study conservation of the folding
nucleus. Instead, they focused on the residues that featured high
$\phi$-values in protein engineering experiments and compared them
with low $\phi$-value residues. As we explained above residues in the
folding nucleus do not necessarily exhibit high $\phi$-values, and
many low $\phi$-value residues are conserved in evolution as they
contribute to stabilization of the native structure. Comparison with
low $\phi$-value residues instead of comparison with the {\em whole}
protein also confused previous analysis since most of $\phi$-values
have been measured for amino acids located in the the core of a
protein and hence these amino acids are on average more conserved.
Here, in contrast, we used the folding nucleus as it was identified
for each protein by the original experimental group and compared its
conservation with the conservation of all amino acids in the protein.

In summary, we showed that folding nucleus is indeed conserved in most
of the proteins whose folding transition states are known from protein
engineering analysis.  That does not mean that folding nucleus
residues are the {\em the only} conserved ones in any family of
homologous proteins. That also may not mean that  folding nucleus is
{\it more} conserved than other residues in the 
 protein core, as nucleus is equally
important for protein stability and for fast folding. Our result show
that the folding nucleus is more conserved than the rest of the
protein. As stated earlier it is difficult to uniquely 
identify folding nucleus by looking at a conservation profile in just
one family of homologous sequences. Nevertheless conservation of
folding nucleus found in this paper and in other works
\cite{EVOL_JMB,LEWYN_NSB} points out to an exciting possibility that
folding rates may be of biological significance. Biological
significance of this fact needs to be assessed in future studies.

\newpage

\pagebreak

\bibliographystyle{apalike}
% \bibliography{../sglit_ES,extra_refs}

\begin{thebibliography}{}

\bibitem[Abagyan \& Batalov, 1997]{Abagyan97}
Abagyan, R. \& Batalov, S. (1997).
\newblock Do aligned sequences share the same fold?
\newblock {\em J Mol Biol}, 273:355--68.

\bibitem[Abkevich {\em et~al.}, 1994]{NUCLEUS}
Abkevich, V., Gutin, A., \& Shakhnovich, E. (1994).
\newblock Specific nucleus as the transition state for protein folding:
  Evidence from the lattice model.
\newblock {\em Biochemistry}, 33:10026--10036.

\bibitem[Bateman {\em et~al.}, 2000]{pfam}
Bateman, A., Birney, E., Durbin, R., Eddy, S., Howe, K., \& Sonnhammer, E.
  (2000).
\newblock The pfam protein families database.
\newblock {\em Nucleic Acids Res}, 28:263--6.

\bibitem[Benner {\em et~al.}, 1994]{GONNET}
Benner, S., Cohen, M., \& Gonnet, G. (1994).
\newblock Amino acid substitution during functionally constrained divergent
  evolution of protein sequences.
\newblock {\em Protein Eng}, 7:1323--32.

\bibitem[Branden \& Tooze, 1998]{BT98}
Branden, C. \& Tooze, J. (1998).
\newblock {\em Introduction to Protein Structure}.
\newblock Garland Publishing, Inc., New York.

\bibitem[Brenner {\em et~al.}, 1998]{Brenner98}
Brenner, S., Chothia, C., \& Hubbard, T. (1998).
\newblock Assessing sequence comparison methods with reliable structurally
  identified distant evolutionary relationships.
\newblock {\em Proc Natl Acad Sci U S A}, 95:6073--8.

\bibitem[Chiti {\em et~al.}, 1999]{Chiti_NSB}
Chiti, F., Taddei, N., White, P., Bucciantini, M., Magherini, F., Stefani, M.,
  \& Dobson, C. (1999).
\newblock Mutational analysis of acylphosphatase suggests the importance of
  topology and contact order in protein folding.
\newblock {\em Nature Structural Biology, in press}, 6:1005--1009.

\bibitem[Dodge {\em et~al.}, 1998]{HSSP}
Dodge, C., Schneider, R., \& Sander, C. (1998).
\newblock The hssp database of protein structure-sequence alignments and family
  profiles.
\newblock {\em Nucleic Acids Res}, 26:313--5.

\bibitem[Dokholyan {\em et~al.}, 2000]{Dokh:2000}
Dokholyan, N., Buldyrev, S., Stanley, H., \& Shakhnovich, E. (2000).
\newblock Identifying the protein folding nucleus using molecular dynamics.
\newblock {\em Journ. Mol. Biol.}, 296:1183--1188.

\bibitem[Fersht, 1997]{FER_COSB}
Fersht, A. (1997).
\newblock Nucleation mechanism of protein folding.
\newblock {\em Curr. Opin. Struct. Biol.}, 7:10--14.

\bibitem[Guo \& Thirumalai, 1995]{GT}
Guo, Z. \& Thirumalai, D. (1995).
\newblock Nucleation mechanism for protein folding and theoretical predictions
  for hydrogen-exchange labelling experiments.
\newblock {\em Biopolymers}, 35:137--139.

\bibitem[Hamill {\em et~al.}, 2000]{Hamill:00}
Hamill, S., Steward, A., \& Clarke, J. (2000).
\newblock The folding of an immunoglobulin-like greek key protein is defined by
  a common-core nucleus and regions constrained by topology.
\newblock {\em J Mol Biol}, 297:165--168.

\bibitem[Henikoff \& Henikoff, 1992]{BLOSUM62}
Henikoff, S. \& Henikoff, J. (1992).
\newblock Amino acid substitution matrices from protein blocks.
\newblock {\em Proc Natl Acad Sci U S A}, 89:10915--9.

\bibitem[Itzhaki {\em et~al.}, 1995]{FER_NUCL}
Itzhaki, L., Otzen, D., \& Fersht, A. (1995).
\newblock The structure of the transition state for folding of chymotrypsin
  inhibitor 2 analyzed by protein engineering methods: Evidence for a
  nucleation-condensation mechanism for protein folding.
\newblock {\em J.Mol.Biol.}, 254:260--288.

\bibitem[Klimov \& Thirumalai, 1998]{KT_JMB}
Klimov, D. \& Thirumalai, D. (1998).
\newblock Lattice models for proteins reveal multiple folding nuclei for
  nucleation-collapse mechanism.
\newblock {\em J.Mol.Biol.}, 282:471--492.

\bibitem[Kragelund {\em et~al.}, 1999]{Poulsen:99}
Kragelund, B., Osmark, P., Neergaard, T., Schiodt, J., Kristiansen, K.,
  Knudsen, J., \& Poulsen, F. (1999).
\newblock The formation of a native-like structure containing eight conserved
  hydrophobic residues is rate limiting in two-state protein folding of acbp.
\newblock {\em Nature Struct Biol}, 6:594--601.

\bibitem[Li {\em et~al.}, 2000]{LEWYN_NSB}
Li, L., Mirny, L., \& Shakhnovich, E. (2000).
\newblock Kinetics, thermodynamics and evolution of non-native interactions in
  protein folding nucleus.
\newblock {\em Nature. Struct. Biol}, 7:336--341.

\bibitem[Lopez-Hernandez \& Serrano, 1996]{CheY}
Lopez-Hernandez, E. \& Serrano, L. (1996).
\newblock Structure of the transition state for folding of the 129 aa protein
  chey resembles that of a smaller protein, ci2.
\newblock {\em Folding \& Design}, 1:43--55.

\bibitem[Lorch {\em et~al.}, 1999]{Lorch99}
Lorch, M., Mason, J., Clarke, A., \& Parker, M. (1999).
\newblock Effects of core mutations on the folding of a beta-sheet protein:
  implications for backbone organization in the i-state.
\newblock {\em Biochemistry}, 38:1377--85.

\bibitem[Main {\em et~al.}, 1999]{Jackson_JMB}
Main, E., Fulton, K., \& Jackson, S. (1999).
\newblock Folding pathway of fkbp12 and characterisation of the transition
  state.
\newblock {\em J. Mol.Biol.}, 291:429--444.

\bibitem[Martinez {\em et~al.}, 1998]{Serrano_NSB1}
Martinez, J., Pissabarro, T., \& Serrano, L. (1998).
\newblock Obligatory steps in protein folding and the conformational diversity
  of the transition state.
\newblock {\em Nature Structural Biology}, 5:721--729.

\bibitem[Michnick \& Shakhnovich, 1998]{UBQ_EVOL}
Michnick, S. \& Shakhnovich, E. (1998).
\newblock A strategy for detecting the conservation of folding-nucleus residues
  in protein superfamilies.
\newblock {\em Folding \& Design}, 3:239--251.

\bibitem[Mirny {\em et~al.}, 1998a]{EVOL2}
Mirny, L., Abkevich, V., \& Shakhnovich, E. (1998a).
\newblock How evolution makes proteins fold quickly.
\newblock {\em Proc Natl. Acad. Sci. USA}, 95:4976--4981.

\bibitem[Mirny {\em et~al.}, 1998b]{PNAS_MAS}
Mirny, L., Abkevich, V., \& Shakhnovich, E. (1998b).
\newblock How evolution makes proteins fold quickly.
\newblock {\em Proc Natl Acad Sci U S A}, 95:4976--81.

\bibitem[Mirny \& EI, ]{MS_REVIEW}
Mirny, L. \& EI, S.
\newblock Protein folding theory: from lattice to all-atom models.
\newblock {\em Annual Review in Biophysics and Biophysical Chemistry}, 30:in
  press.

\bibitem[Mirny \& Shakhnovich, 1999]{EVOL_JMB}
Mirny, L. \& Shakhnovich, E. (1999).
\newblock Universally conserved residues in protein folds. reading evolutionary
  signals about protein function, stability and folding kinetics.
\newblock {\em J.Mol.Biol.}, 291:177--196.

\bibitem[Pande {\em et~al.}, 1998]{CO_PANDE}
Pande, V., Grosberg, A., Rokshar, D., \& Tanaka, T. (1998).
\newblock Pathways for protein folding: is a ``new view'' needed?
\newblock {\em Curr Opin Struct Biology}, 8:68--79.

\bibitem[Plaxco {\em et~al.}, 2000]{Plaxco:2000}
Plaxco, K., Larson, S., Ruczinski, I., Riddle, D., Buchwitz, B., Davidson, A.,
  \& Baker, D. (2000).
\newblock Evolutionary conservation in protein folding kinetics.
\newblock {\em J. Mol.Biol.}, 298:303--312.

\bibitem[Ptitsyn, 1998]{PTIT_EV}
Ptitsyn, O. (1998).
\newblock Protein folding and protein evolution: Common folding nucleus in
  different subfamilies of c-type cytochromes?
\newblock {\em J.Mol.Biol}, 278:655--666.

\bibitem[Ptitsyn \& Ting, 1999]{PTIT_EV2}
Ptitsyn, O. \& Ting, K. (1999).
\newblock Non-functional conserved residues in globins and their possible role
  as a folding nucleus.
\newblock {\em J.Mol.Biol}, 291:671--682.

\bibitem[Schneider, 1999]{Schneider99}
Schneider, T. (1999).
\newblock Measuring molecular information [letter].
\newblock {\em J Theor Biol}, 201:87--92.

\bibitem[Shakhnovich, 1997]{COSB}
Shakhnovich, E. (1997).
\newblock Theoretical studies of protein-folding thermodynamics and kinetics.
\newblock {\em Curr. Opin. Struct. Biol.}, 7:29--40.

\bibitem[Shakhnovich {\em et~al.}, 1996]{ALAMUT}
Shakhnovich, E., Abkevich, V., \& Ptitsyn, O. (1996).
\newblock Conserved residues and the mechanism of protein folding.
\newblock {\em Nature}, 379:96--98.

\bibitem[Stormo, 1998]{Stormo98}
Stormo, G. (1998).
\newblock Information content and free energy in dna--protein interactions.
\newblock {\em J Theor Biol}, 195:135--7.

\bibitem[Ternstrom {\em et~al.}, 1999]{Oliveberg_U1A}
Ternstrom, T., Mayor, U., Akke, M., \& Oliveberg, M. (1999).
\newblock From snap-shot to movie: phi-value analysis of protein folding
  transition states taken one step further.
\newblock {\em Proc Natl Acad Sci USA}, 96:14854--14859.

\bibitem[Thompson \& Goldstein, 1996]{Thompson96}
Thompson, M. \& Goldstein, R. (1996).
\newblock Constructing amino acid residue substitution classes maximally
  indicative.
\newblock {\em Proteins}, 25:28--37.

\bibitem[Vilegas {\em et~al.}, 1998]{Serrano_ADA2H}
Vilegas, V., Martinez, J., Avilez, F., \& Serrano, L. (1998).
\newblock Structure of the transition state in the folding process of human
  procarboxypeptidase a2 activation domain.
\newblock {\em J Mol Biol.}, 283:1027--1036.

\end{thebibliography}

\pagebreak

\section*{Figure Captions}

{\bf Fig.1} Variability profiles (sequence entropy) for nine different
proteins computed using MS residue classes. Circles indicate positions
at which $\phi$-values have been experimentally measured. Residues
forming the folding nucleus are shown by filled circles.

{\bf Fig.2} Nine studied proteins with C$_\beta$ atoms colored according to
the degree of their conservation (evaluated in Fig.1): from 
blue (high conservation) to light-blue,
green, yellow and red (no conservation). Folding 
nucleus residues are shown 
by twice
as large spheres. Notice conserved (blue) cores 
of the proteins and
non-conserved (yellow and red) surfaces. Also 
notice several conserved
non-nucleus residues in the protein core.

\pagebreak

\setlength{\topmargin}{-1.0in}
\setlength{\textheight}{9.5in}
 
\pagestyle{empty}

\begin{figure}[htb]
\epsfysize 8.5 in
\centerline{\epsfbox{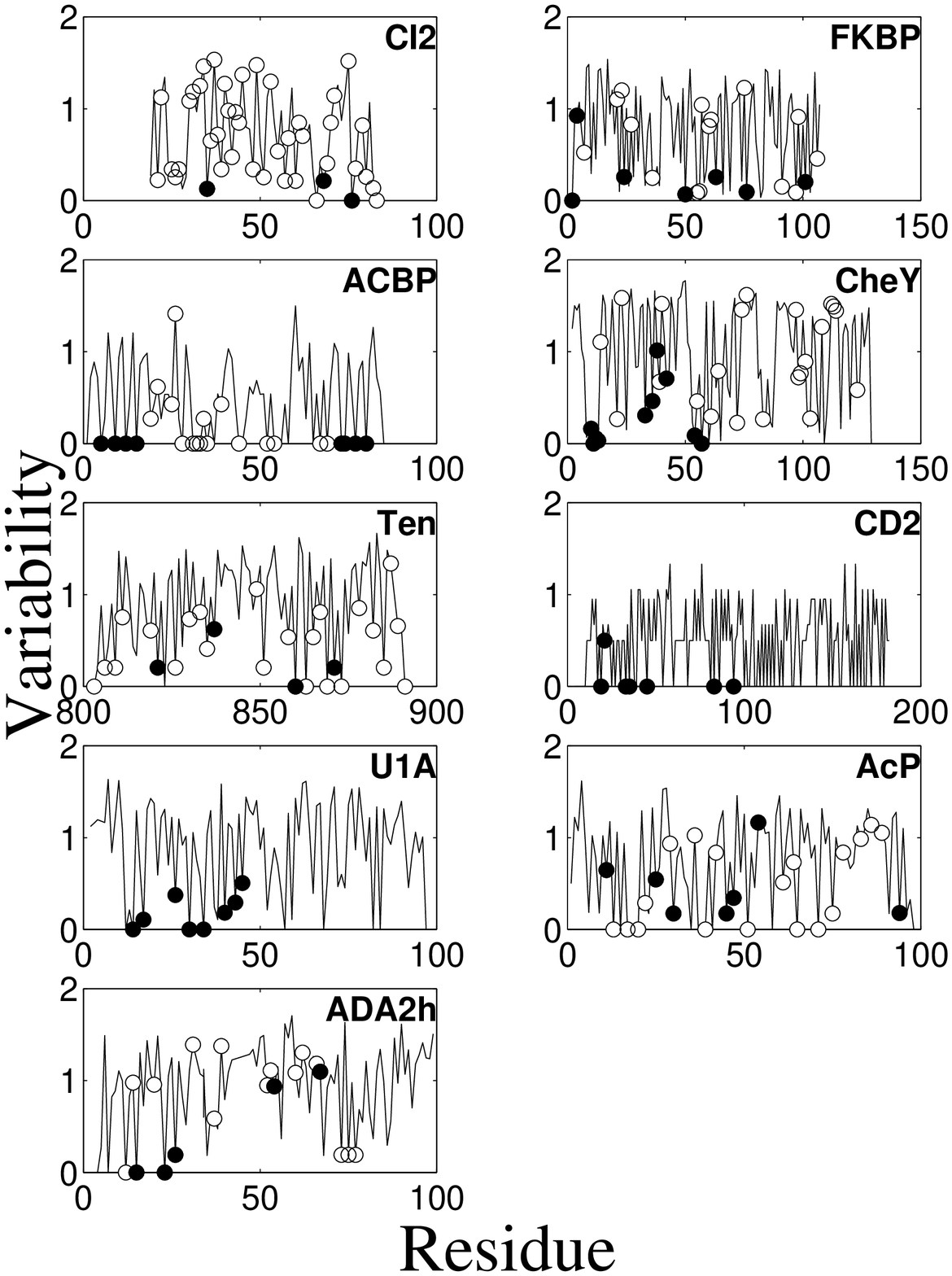}}
\caption{}
\label{fig:conserv_all}
\end{figure}

\pagebreak
\pagestyle{empty}

\begin{figure}[htb]
\epsfysize 8.5 in
\centerline{\epsfbox{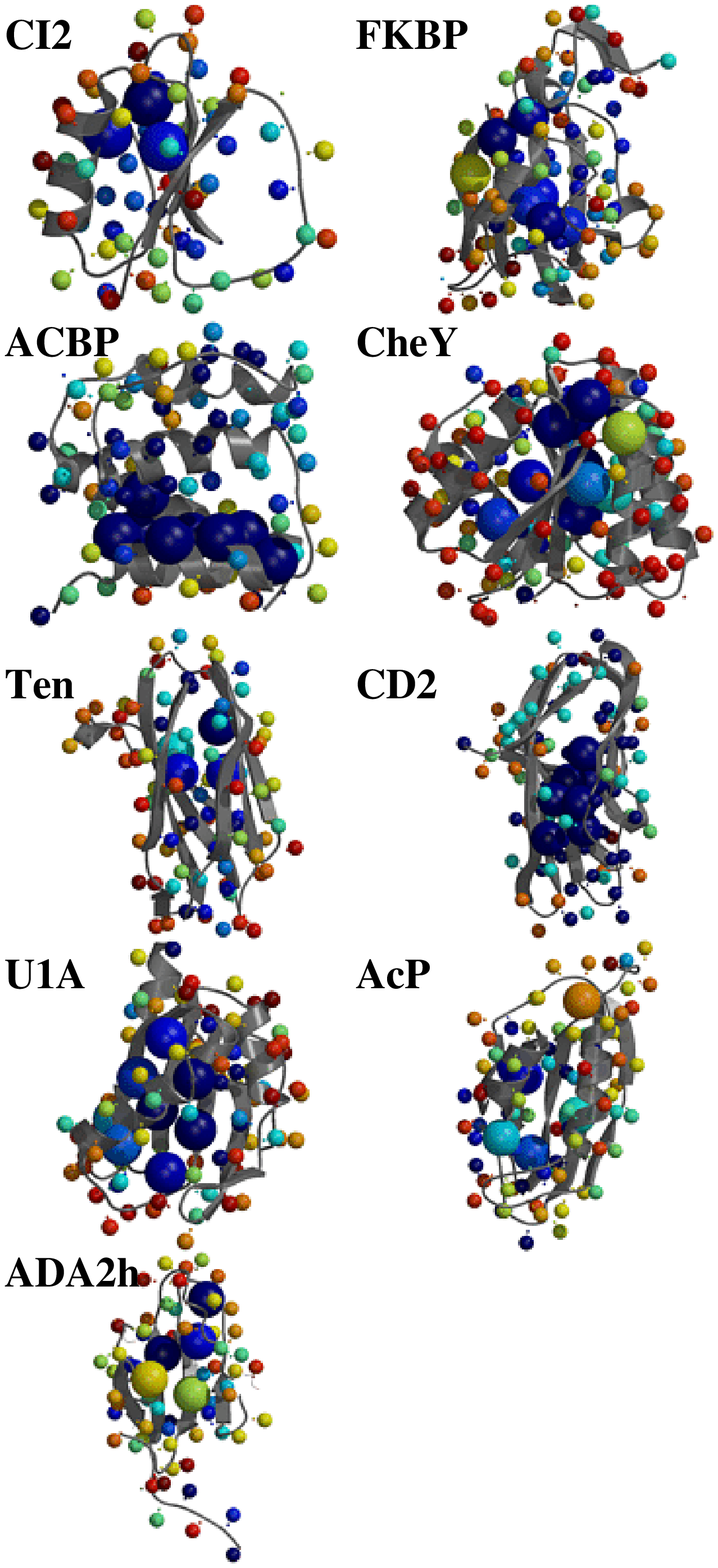}}
\caption{}
\label{fig:structures}
\end{figure}

\pagebreak
\setlength{\oddsidemargin}{-0.7in}
\setlength{\evensidemargin}{-0.7in}
\setlength{\textwidth}{7.5in}

\begin{table}
\begin{tabular}{|l|l|l|l|}
\hline
\bf Protein & \bf PDB & \bf Folding Nucleus & \bf Reference \\ \hline
CI2	& 2ci2I	& A35 L68 I76			& \cite{FER_NUCL} \\ \hline
Tenascin & 1ten	& I821 Y837 I860 V871 		& \cite{Hamill:00} \\ \hline
CD2.d1	& 1hnf & L19 I21 I33 A45 V83 L94    W35 & \cite{Lorch99} \\ \hline
CheY	& 3chy & D12 D13 D57   V10 V11 V33 A36     D38    A42 V54 &
\cite{CheY} \\ \hline
ADA2h	& 1aye & I15 L26 F67 V54 I23 		& \cite{Serrano_ADA2H} \\ \hline
AcP	& 1aps, 2acy & Y11 P54 F94		& \cite{Chiti_NSB} \\ \hline
U1A	& 1urn & I43 V45     L30 F34 I40  I14 L17 L26 & \cite{Oliveberg_U1A} \\
\hline
ACBP	& 1aca & F5 A9 V12 L15 Y73 I74 V77 L80		& \cite{Poulsen:99} \\ \hline
FKBP12	& 1fkj & V2 V4 V24 V63 I76 I101  	& \cite{Jackson_JMB} \\ \hline
% SH3 $\alpha$-spec 	&	& 		& 		\\ \hline
% SH3 src			&	& 		&		\\
\end{tabular}
\caption{Folding nuclei as identified by the authors}
\end{table}

\pagebreak
\newpage 

\begin{table}
\begin{tabular}{|l|l||l||l|l|l|l|l|l|}
\hline
         & MS & BT & no grouping  & [I,L,V], [W,F,Y]      & [I,L,V]        & [I,L,V]      & [I,L,V] \\ 
         &    &    &              &          [R,K] [D,E]  &        [W,F,Y] &        [W,F] &         \\ \hline 
 $N_{class}$  &  6  &  5  &  20  &  14  &  16  &  17  &  18  \\ \hline 
 CI2      &  \bf 0.0041 &  $0.01$   &  $0.0382$ &  $0.007$  &  $0.002$  &  $0.004$  &  $0.0044$   \\ \hline 
 FKBP12    &  \bf 0.0187  &  $0.02$    &  $0.1585$  &  $0.044$   &  $0.047$   &  $0.053$   &  $0.0363$    \\ \hline 
 ACBP        &  $\bf <10^{-5}$  &  $<10^{-5}$  &  $0.0216$    &  $0.022$     &  $0.008$     &  $0.0080$    &  $0.0067$      \\ \hline 
 CheY        &  $\bf <10^{-5}$  &  $<10^{-5}$  &  $0.0011$    &  $0.0040$    &  $0.0050$    &  $0.0020$    &  $0.0022$      \\ \hline 
 Ten      &  $\bf 0.008$  &  $0.018$  &  $0.2477$ &  $0.0260$ &  $0.0220$ &  $0.0130$ &  $0.0197$   \\ \hline 
 CD2.d1      &  $\bf <10^{-5}$  &  $<10^{-5}$  &  $<10^{-5}$  &  $<10^{-3}$  &   $<10^{-3}$ &   $<10^{-3}$  &  $<10^{-5}$    \\ \hline 
 U1A       &  $\bf 0.0009$  &  $0.001$   &  $0.0029$  &  $<10^{-3}$       &  $<10^{-3}$  &  $<10^{-3}$  &  $0.0002$    \\ \hline 
 AcP      &  $\bf 0.089$  &  $0.086$  &  $0.0126$ &  $0.025$  &  $0.021$  &  $0.009$  &  $0.0136$   \\ \hline 

\end{tabular}   
\caption{Probability $P_0$ of nucleus being as conserved as
the whole protein (see text for details) computed for all nine
proteins and seven different classification schemes. MS as in
\cite{PNAS_MAS,EVOL_JMB}, BT as in \cite{BT98}: hydrophobic [A V F P M
I L],polar [S T Y H C N Q W],basic [R K],acidic [D E],gly [G]),
$N_{class}$ - number of groups in
each classification}
\end{table}

\end{document}